# New expression on the performance of a novel multi-hop relay-assisted hybrid FSO / RF communication system

Mohammad Ali Amirabadi

*Abstract*— In this paper a novel multi-hop relay-assisted hybrid FSO / RF system is presented. It is assumed that direct RF connection between mobile users and source Base Station is impossible, therefore a relay connects RF users to the source Base Station via a FSO link. Source and destination Base Stations are connected via a multi-hop relay-assisted hybrid FSO / RF link. FSO link is investigated over moderate to saturate regimes of atmospheric turbulence. Also RF link is assumed to have Rayleigh fading. For the first time, new exact and asymptotic expressions are derived in closed-form for Bit Error Rate (BER) and the Outage Probability ($P_{out}$), of the proposed system. MATLAB simulations are performed to validate obtained analytical results. Results indicate that the proposed structure has low dependence on the number of users, therefore, the proposed structure is suitable for cells which encounter different populations during a day, because there is little performance difference between systems with different number of users. Also the proposed structure, at Negative exponential atmospheric turbulence has small dependence on the number of relays, but this dependence is a bit more for Gamma-Gamma atmospheric turbulence. Therefore, the proposed structure increases capacity whereas maintaining performance of the system.

## I. INTRODUCTION

During the last decades, FSO communication system is taken into considerations due to the advances in optoelectronic devices and reduction of their cost. FSO system has large bandwidth and therefore is suitable for the last-mile backup application of 4G and 5G communication systems. This system is highly secured and has simple and chip setup, and especially because of unlimited and license-free bandwidth, is suitable for bottleneck applications [1].

One of the main problems of FSO system, is its high sensitivity to weather conditions. But even at clear weather, this system encounters with the problem of atmospheric turbulence, caused by heterogeneous distribution of temperature and pressure of the atmosphere. Atmospheric turbulence causes atmospheric scintillation, this effect is the same as RF system fading and is accompanied by random fluctuations of the received signal intensity[2]. FSO transreceiver is deployed at the top of high buildings; light earthquakes and winds cause misalignment of FSO transreceiver. This effect that is called pointing error, significantly degrades performance of FSO system. Various statistical distributions have been used in order to investigate the effects of atmospheric turbulences; such as Exponential-Weibull [3], Generalized Malaga,[4] Log-normal [5], Gamma-Gamma [6], and Negative Exponential [7]. Gamma-Gamma and Negative exponential have high accompany with experimental results, respectively for moderate to strong and saturate regimes of atmospheric turbulences.

During the last decades several studies have been done on the topic of FSO cooperative communication systems; cooperative communication is an efficient way of improving system performance without so much additional hardware requirement. The main difference of cooperative communication systems is related to their process on the received signal. Various protocols have been studied for FSO cooperative communication systems, including decode and forward [8], amplify and forward [9], quantize and forward[10]. In decode and forward protocol, relay decodes the received signal and then forwards it. In order to have more error control, the decoded signal can also be re-encoded and forwarded to the destination. In decode and forward protocol in many applications, because of some limitations of transceiver or lack of Channel Codebook Information, channel coding may not be desirable. In this situations, the relay detects or demodulates the

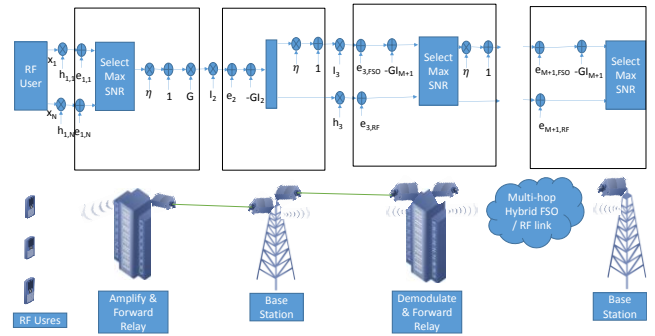

Fig. 1. The proposed multi-hop relay-assisted hybrid FSO / RF system.

M. A. Amirabadi is with the School of Electrical Engineering, Iran University of Science and Technology (IUST), Tehran 1684613114, Iran(email: m_amirabadi@elec.iust.ac.ir)

received signal and then forwards it [10].

Although that RF communication systems have stable channel, but they have low data rate and sometimes are limited by rain. One way to achieve a reliable and accessible communication is to combine FSO and RF systems. These systems, are called hybrid FSO / RF and even at bad weather conditions offer high link reliability and accessibility. Hybrid FSO / RF links are implemented in both series [10] and parallel [12] structures. In parallel structure data is transmitted via FSO and RF links simultaneously [13] or by use of a switch [14]. In this paper the link between source and destination Base Stations use simultaneous data transmission.

Many works have been done about hybrid FSO / RF systems, they have mostly investigated one hop [14] and two hop [16] structures, few papers investigated multihop structures with series [17] and parallel [18] hops. In this paper, a novel multi-hop relay-assisted hybrid FSO / RF system is presented. In this system, a relay connects mobile user to the source Base Station via a FSO link and a multi-hop relay-assisted hybrid FSO / RF link connects source and destination Base Stations. This structure is recommended for places where direct RF connection between mobile users and source Base Station is impossible, such as impassable areas and places where RF link is disrupted by heavy rain. Mostly, studies done about hybrid FSO / RF systems, investigate the effects of various channel and system model; it is the first time that a multiuser scheme is presented for a multi-hop hybrid FSO / RF structure. The first relay, between different mobile users, communicates with the one with the highest SNR at the relay input. However, the relay amplifies and forwards the selected signal. Both cases of known Channel State Information (CSI) and unknown CSI at the first relay, are investigated. It is assumed that the multi-hop link between source and destination Base Stations, use demodulate and forward protocol. For the first time, such a structure is investigated in wide range of atmospheric turbulences, form moderate to saturate regimes, by Gamma-Gamma and Negative Exponential models, respectively; also in moderate to strong regimes, the effect of pointing error is also considered to accompany with actual results. It is the first time that the effect of pointing error is considered in a multi-hop hybrid FSO / RF structure. For the first time, new exact and asymptotic expressions are derived in closed-form for BER and $P_{out}$ of the proposed structure. Derived expressions are validated through MATLAB simulations. Multihop structure of the proposed system significantly increases capacity while reducing total consumed power; also simultaneous parallel data transmission, significantly improves system performance, because there is almost no condition which degrades both FSO and RF systems.

The remainder of the paper is organized as follows: system model is expressed in Section II, the proposed structure is investigated in cases of known CSI and unknown CSI in sections III and IV, respectively. In section V simulation results are compared with exact and asymptotic analytical results. Section VI is the conclusions of this study.

## II. SYSTEM MODEL

As can be seen in Fig. 1, in this structure there is a relay between mobile users and the source Base Station. Relay between received RF signals from users, selects the one with the highest SNR. Assuming $x_i$ as the transmitted signal from $i-th$ user, this signal is broadcasted in Rayleigh channel with fading coefficient of $h_{1,i}$, and at the relay receiver input is added by $e_{1,i}$, the Additive White Gaussian Noise (AWGN) with zero mean and $\sigma_{RF}^2$ variance. The received signal at the first relay is as follows:

$$y_{1,i} = x_i h_{1,i} + e_{1,i} \qquad (1)$$

Assuming $y_1$ as the selected RF signal; it is converted to FSO signal by conversion efficiency of $\eta$, then is added by a DC signal with unit amplitude in order to be positive. FSO signal is then amplified and forwarded at the second hop, as follows:

$$x_2 = G(1 + \eta y_1) \qquad (2)$$

This signal is transmitted through FSO channel with atmospheric turbulence intensity of $I_2$ and at the source Base Station, is added by $e_2$, AWGN with zero mean and $\sigma_{FSO}^2$ variance. After DC removal, the received signal at the second relay (source Base Station) becomes as follows:

$$y_2 = G\eta I_2 h_1 x + G\eta I_2 e_1 + e_2 \qquad (3)$$

Instantaneous SNR at the second relay input is as follows:

$$\gamma_{2^{nd}relay} = \frac{G^2 \eta^2 I_2^2 h_1^2 E[x^2]}{G^2 \eta^2 I_2^2 \sigma_{RF}^2 + \sigma_{FSO}^2} \qquad (4)$$

In the case of known CSI, the first relay amplifies the signal with gain of $G^2 = 1/(h_1^2 + \sigma_{RF}^2)$ [19]. Substituting $\gamma_1 = h_1^2/\sigma_{RF}^2, \gamma_2 = G^2 \eta^2 I_2^2/\sigma_{FSO}^2$ and $G$, (4) becomes equal to:

$$\gamma_{2^{nd}relay} = \frac{\gamma_1 \gamma_2}{\gamma_1 + \gamma_2 + 1} \qquad (5)$$

In the case of unknown CSI, the first relay amplifies the signal with gain of $G^2 = 1/(C\sigma_{RF}^2)$ [19]. Substituting $\gamma_1 = h_1^2/\sigma_{RF}^2, \gamma_2 = G^2 \eta^2 I_2^2/\sigma_{FSO}^2$ and $G$, (4) becomes equal to:

$$\gamma_{2^{nd}relay} = \frac{\gamma_1 \gamma_2}{C + \gamma_2} \qquad (6)$$

The second relay demodulates received signal, then generates two copies of it, then modulates and forwards them at FSO and RF links. Assuming $M$ as the number of relays, at $j-th; j = 3,4,...,M+1$ hop, FSO and RF signals encounter with atmospheric turbulence with $I_j$ intensity fading with $h_j$ coefficient, respectively. Received FSO and RF signals at the $j-th$ receiver input are added by AWGN with zero mean and

$\sigma_{FSO}^2$ and $\sigma_{RF}^2$ variances, then between them the one with higher SNR is selected and demodulated, then modulated and forwarded in FSO and RF links. The same procedure is continued till the destination Base Station.

The pdf and CDF of Gamma-Gamma distribution with the effect of pointing error[14] and the CDFs of Rayleigh and Negative Exponential distribution are respectively as follows:

$$f_\gamma(\gamma) = \frac{\xi^2}{2\Gamma(\alpha)\Gamma(\beta)\gamma} G_{1,4}^{3,0}\left(\alpha\beta\kappa\sqrt{\frac{\gamma}{\overline{\gamma}_{FSO}}} \bigg| \begin{array}{c} \xi^2+1 \\ \xi^2, \alpha, \beta \end{array}\right) \quad (7)$$

$$F_\gamma(\gamma) = \frac{\xi^2}{\Gamma(\alpha)\Gamma(\beta)} G_{2,4}^{3,1}\left(\alpha\beta\kappa\sqrt{\frac{\gamma}{\overline{\gamma}_{FSO}}} \bigg| \begin{array}{c} 1, \xi^2+1 \\ \xi^2, \alpha, \beta, 0 \end{array}\right) \quad (8)$$

$$F_\gamma(\gamma) = 1 - e^{-\frac{\gamma}{\overline{\gamma}_{RF}}} \quad (9)$$

$$F_\gamma(\gamma) = 1 - e^{-\lambda\sqrt{\frac{\gamma}{\overline{\gamma}_{FSO}}}} \quad (10)$$

Since at the first relay the signal with the highest SNR is selected for communication, the Cumulative Density Function (CDF) of $\gamma_1$, the instantaneous SNR at the first relay input, becomes as follows:

$$F_{\gamma_1}(\gamma) = \Pr(\max(\gamma_{1,1}, \gamma_{1,2}, \dots, \gamma_{1,N}) \leq \gamma) = \Pr(\gamma_{1,1} \leq \gamma, \gamma_{1,2} \leq \gamma, \dots, \gamma_{1,N} \leq \gamma) \quad (11)$$

where $\gamma_{1,i}$ is instantaneous SNR at the first relay input of $i-th$ path. Assuming independent identically distributed different RF paths, and using (9), the CDF of instantaneous SNR at the first relay input becomes as follows:

$$F_{\gamma_1}(\gamma) = \prod_{i=1}^{N} F_{\gamma_{1,i}}(\gamma) = \left(F_{\gamma_{1,i}}(\gamma)\right)^N = \left(1 - e^{-\frac{\gamma}{\overline{\gamma}_{RF}}}\right)^N \quad (12)$$

Differentiating of (12), the probability distribution function (pdf) of instantaneous SNR at the first relay input becomes:

$$f_{\gamma_1}(\gamma) = \frac{N}{\overline{\gamma}_{RF}} e^{-\frac{\gamma}{\overline{\gamma}_{RF}}} \left(1 - e^{-\frac{\gamma}{\overline{\gamma}_{RF}}}\right)^{N-1} = \sum_{k=0}^{N-1} \binom{N-1}{k} (-1)^k \frac{N}{\overline{\gamma}_{RF}} e^{-\frac{(k+1)\gamma}{\overline{\gamma}_{RF}}} \quad (13)$$

Since at the $j-th$ relay, the signal with higher SNR is selected for communication, and assuming independent FSO and RF links, the CDF of $\gamma_j$, the instantaneous SNR at $j-th$ relay input, becomes as follows:

$$F_{\gamma_j}(\gamma) = \Pr(\max(\gamma_{\gamma_{FSO},j}, \gamma_{\gamma_{RF},j}) \leq \gamma) = \Pr(\gamma_{\gamma_{FSO},j} \leq \gamma, \gamma_{\gamma_{RF},j} \leq \gamma) = F_{\gamma_{FSO},j}(\gamma) F_{\gamma_{RF},j}(\gamma) \quad (14)$$

III. PERFORMANCE EVALUATION OF KNOWN CSI AT THE FIRST RELAY SCHEME

Assuming high SNR approximation (5) becomes equal to:

$$\gamma_{2nd_{relay}} = \frac{\gamma_1 \gamma_2}{\gamma_1 + \gamma_2 + 1} \cong \min(\gamma_1, \gamma_2) \quad (15)$$

Therefore the CDF of $\gamma_{2nd_{relay}}$ random variable becomes:

$$F_{\gamma_{2nd_{relay}}}(\gamma) = \Pr(\gamma_{2nd_{relay}} \leq \gamma) \quad (16)$$
$$= 1 - \Pr(\min(\gamma_1, \gamma_2) \geq \gamma)$$
$$= 1 - \Pr(\gamma_1 \geq \gamma)\Pr(\gamma_2 \geq \gamma)$$
$$= 1 - (1 - F_{\gamma_1}(\gamma))(1 - F_{\gamma_2}(\gamma))$$

Substituting (8) and (12) into (16), and by using binomial expansion theorem, the CDF of $\gamma_{2nd_{relay}}$ in Gamma-Gamma atmospheric turbulence with the effect of pointing error becomes as follows:

$$F_{\gamma_{2nd_{relay}}}(\gamma) = 1 + \sum_{k=1}^{N} \binom{N}{k}(-1)^k e^{-\frac{k\gamma}{\overline{\gamma}_{RF}}} \left(1 - \frac{\xi^2}{\Gamma(\alpha)\Gamma(\beta)} G_{2,4}^{3,1}\left(\alpha\beta\kappa\sqrt{\frac{\gamma}{\overline{\gamma}_{FSO}}} \bigg| \begin{array}{c} 1, \xi^2+1 \\ \xi^2, \alpha, \beta, 0 \end{array}\right)\right) \quad (17)$$

Substituting (10) and (12) into (16), and using binomial expansion theorem, the CDF of $\gamma_{2nd_{relay}}$ in Negative Exponential atmospheric turbulence becomes as follows:

$$F_{\gamma_{2nd_{relay}}}(\gamma) = 1 + \sum_{k=1}^{N} \binom{N}{k}(-1)^k e^{-\frac{k\gamma}{\overline{\gamma}_{RF}}} e^{-\lambda\sqrt{\frac{\gamma}{\overline{\gamma}_{FSO}}}} \quad (18)$$

A. Outage Probability

Assuming that error occurs only because of wrong demodulation at each relay, and according to $P_{out}(\gamma_{th}) = F_\gamma(\gamma_{th})$, $P_{out}$ of the proposed structure becomes equal to:

$$P_{out}(\gamma_{th}) = \Pr\{(\gamma_1, \gamma_2, \dots, \gamma_{M+1}) \leq \gamma_{th}\} = 1 - \Pr\{\gamma_1 \geq \gamma_{th}, \gamma_2 \geq \gamma_{th}, \dots, \gamma_{M+1} \geq \gamma_{th}\} = 1 - (1 - \Pr\{(\gamma_1, \gamma_2) \leq \gamma_{th}\})(1 - \Pr\{(\gamma_3 \leq \gamma_{th})\}) \dots (1 - \Pr\{(\gamma_{M+1} \leq \gamma_{th})\}) = 1 - \left(1 - F_{\gamma_{2nd_{relay}}}(\gamma_{th})\right)\left(1 - F_{\gamma_3}(\gamma_{th})\right) \dots \left(1 - F_{\gamma_{M+1}}(\gamma_{th})\right) \quad (19)$$

According to (14), after substituting (17), (8) and (9) into (19), $P_{out}$ of the proposed structure in Gamma-Gamma atmospheric turbulence with the effect of pointing error is equal to:

$$P_{out}(\gamma_{th}) = 1 + \sum_{k=1}^{N} \binom{N}{k}(-1)^k e^{-\frac{k\gamma_{th}}{\overline{\gamma}_{RF}}} \left(1 - \frac{\xi^2}{\Gamma(\alpha)\Gamma(\beta)} G_{2,4}^{3,1}\left(\alpha\beta\kappa\sqrt{\frac{\gamma_{th}}{\overline{\gamma}_{FSO}}} \bigg| \begin{array}{c} 1, \xi^2+1 \\ \xi^2, \alpha, \beta, 0 \end{array}\right)\right) \left[1 - \frac{\xi^2}{\Gamma(\alpha)\Gamma(\beta)}\left(1 - e^{-\frac{\gamma_{th}}{\overline{\gamma}_{RF}}}\right) G_{2,4}^{3,1}\left(\alpha\beta\kappa\sqrt{\frac{\gamma_{th}}{\overline{\gamma}_{FSO}}} \bigg| \begin{array}{c} 1, \xi^2+1 \\ \xi^2, \alpha, \beta, 0 \end{array}\right)\right]^{M-1} \quad (20)$$

$$P_e = \frac{1}{2}\left\{1 + \sum_{k=1}^{N}\sum_{t=0}^{M-1}\sum_{u=0}^{t}\sum_{k_1=0}^{t}\sum_{k_2=0}^{k_1}\sum_{n=0}^{\infty} \Omega \binom{t}{k_1}\binom{k_1}{k_2} X_0^{t-k_1}\left(Y_n^{(k_1-k_2)} * Z_n^{(k_2)}\right) \frac{\left(\frac{1}{\bar{\gamma}_{FSO}}\right)^{\frac{n+\xi^2(t-k_1)+\alpha(k_1-k_2)+\beta k_2}{2}}}{\left(1+\frac{(k+u)}{\bar{\gamma}_{RF}}\right)^{1+\frac{n+\xi^2(t-k_1)+\alpha(k_1-k_2)+\beta k_2}{2}}} \times\right.$$

$$\left.\left[\Gamma\left(1 + \frac{n+\xi^2(t-k_1)+\alpha(k_1-k_2)+\beta k_2}{2}\right) - \frac{\xi^2 2^{\alpha+\beta-3}}{\pi\Gamma(\alpha)\Gamma(\beta)} \times G_{5,8}^{6,3}\left(\frac{(\alpha\beta\kappa)^2}{16\bar{\gamma}_{FSO}\left(1+\frac{(k+u)}{\bar{\gamma}_{RF}}\right)}\middle|\begin{array}{c}-\frac{n+\xi^2(t-k_1)+\alpha(k_1-k_2)+\beta k_2}{2}, \frac{1}{2}, 1, \frac{\xi^2+1}{2}, \frac{\xi^2+2}{2}\\ \frac{\xi^2}{2}, \frac{\xi^2+1}{2}, \frac{\alpha}{2}, \frac{\alpha+1}{2}, \frac{\beta}{2}, \frac{\beta+1}{2}, 0, \frac{1}{2}\end{array}\right)\right]\right\}$$
(26)

Substituting binomial expansion of $\left[1 - \frac{\xi^2}{\Gamma(\alpha)\Gamma(\beta)}\left(1 - e^{-\frac{\gamma_{th}}{\bar{\gamma}_{RF}}}\right) G_{2,4}^{3,1}\left(\alpha\beta\kappa\sqrt{\frac{\gamma_{th}}{\bar{\gamma}_{FSO}}}\middle|\begin{array}{c}1,\xi^2+1\\ \xi^2,\alpha,\beta,0\end{array}\right)\right]^{M-1}$,

as $\sum_{t=0}^{M-1}\sum_{u=0}^{t}\binom{M-1}{t}\binom{t}{u}(-1)^{t+u}e^{-\frac{u\gamma_{th}}{\bar{\gamma}_{RF}}}\left(\frac{\xi^2}{\Gamma(\alpha)\Gamma(\beta)} \times G_{2,4}^{3,1}\left(\alpha\beta\kappa\sqrt{\frac{\gamma_{th}}{\bar{\gamma}_{FSO}}}\middle|\begin{array}{c}1,\xi^2+1\\ \xi^2,\alpha,\beta,0\end{array}\right)\right)^t$, $P_{out}$ of the proposed structure in Gamma-Gamma atmospheric turbulence with the effect of pointing errors becomes equal to:

$$P_{out}(\gamma_{th}) = 1 + \sum_{k=1}^{N}\sum_{t=0}^{M-1}\sum_{u=0}^{t}\Omega \times \quad (21)$$
$$e^{-\frac{(k+u)\gamma_{th}}{\bar{\gamma}_{RF}}}\left(\frac{\xi^2}{\Gamma(\alpha)\Gamma(\beta)} G_{2,4}^{3,1}\left(\alpha\beta\kappa\sqrt{\frac{\gamma_{th}}{\bar{\gamma}_{FSO}}}\middle|\begin{array}{c}1,\xi^2+1\\ \xi^2,\alpha,\beta,0\end{array}\right)\right)^t \times$$
$$\left[1 - \frac{\xi^2}{\Gamma(\alpha)\Gamma(\beta)} G_{2,4}^{3,1}\left(\alpha\beta\kappa\sqrt{\frac{\gamma_{th}}{\bar{\gamma}_{FSO}}}\middle|\begin{array}{c}1,\xi^2+1\\ \xi^2,\alpha,\beta,0\end{array}\right)\right]$$

where $\Omega = \binom{N}{k}\binom{M-1}{t}\binom{t}{u}(-1)^{k+t+u}$.

According to (14) and substituting (18), (9) and (10) into (19), $P_{out}$ of the proposed structure for Negative Exponential atmospheric turbulence is equal to:

$$P_{out}(\gamma_{th}) = 1 + \sum_{k=1}^{N}\binom{N}{k}(-1)^k e^{-\frac{k\gamma_{th}}{\bar{\gamma}_{RF}}} \times \quad (22)$$
$$e^{-\lambda\sqrt{\frac{\gamma_{th}}{\bar{\gamma}_{FSO}}}}\left[1 - \left(1 - e^{-\frac{\gamma_{th}}{\bar{\gamma}_{RF}}}\right)\left(1 - e^{-\lambda\sqrt{\frac{\gamma_{th}}{\bar{\gamma}_{FSO}}}}\right)\right]^{M-1}$$

Substituting binomial expansion of $\left[1 - \left(1 - e^{-\gamma_{th}/\bar{\gamma}_{RF}}\right) \times \left(1 - e^{-\lambda\sqrt{\gamma_{th}/\bar{\gamma}_{FSO}}}\right)\right]^{M-1}$, as $\sum_{t=0}^{M-1}\sum_{u=0}^{t}\sum_{v=0}^{t}\binom{M-1}{t}\binom{t}{u}\binom{t}{v}(-1)^{t+u+v}e^{-u\gamma_{th}/\bar{\gamma}_{RF}}e^{-\lambda v\sqrt{\gamma_{th}/\bar{\gamma}_{FSO}}}$, and substituting equivalent Meijer-G form of $e^{-\lambda(v+1)\sqrt{\gamma_{th}/\bar{\gamma}_{FSO}}}$ as $\frac{1}{\sqrt{\pi}}G_{0,2}^{2,0}\left(\frac{(\lambda(v+1))^2\gamma_{th}}{4\bar{\gamma}_{FSO}}\middle|\begin{array}{c}-\\ 0,0.5\end{array}\right)$ [19], $P_{out}$ of the proposed structure in Negative Exponential atmospheric turbulence becomes equal to:

$$P_{out}(\gamma_{th}) = 1 + \sum_{k=1}^{N}\sum_{t=0}^{M-1}\sum_{u=0}^{t}\sum_{v=0}^{t}\Lambda\, e^{-\frac{(k+u)\gamma_{th}}{\bar{\gamma}_{RF}}} \quad (23)$$
$$\times G_{0,2}^{2,0}\left(\frac{(\lambda(v+1))^2\gamma_{th}}{4\bar{\gamma}_{FSO}}\middle|\begin{array}{c}-\\ 0,0.5\end{array}\right)$$

where $\Lambda = \binom{N}{k}\binom{M-1}{t}\binom{t}{u}\binom{t}{v}(-1)^{k+t+u+v}\frac{1}{\sqrt{\pi}}$.

### B. Bit Error Rate

Differential modulations are less sensitive to noise and interference, and because of the following reasons, their detection is optimal: no need for CSI or complex processing at the receiver, no need for threshold adjustment feedback, no effect on system throughput, due to lack of pilot or training sequence, reducing the effects of background noise at the receiver, reducing the effects of pointing errors and whether conditions such as fog and mist [21]. BER of DPSK modulation is calculated from the following[13]:

(24)
$$P_e = \frac{1}{2}\int_0^{\infty} e^{-\gamma}F_\gamma(\gamma)d\gamma = \frac{1}{2}\int_0^{\infty} e^{-\gamma}P_{out}(\gamma)d\gamma$$

Substituting (21) into (24), BER of DPSK modulation in Gamma-Gamma atmospheric turbulence with the effect of pointing error is equal to:

$$P_e = \frac{1}{2}\int_0^{\infty} e^{-\gamma}\left\{1 + \sum_{k=1}^{N}\sum_{t=0}^{M-1}\sum_{u=0}^{t}\Omega e^{-\frac{(k+u)\gamma}{\bar{\gamma}_{RF}}} \times \right. \quad (25)$$
$$\left(\frac{\xi^2}{\Gamma(\alpha)\Gamma(\beta)} G_{2,4}^{3,1}\left(\alpha\beta\kappa\sqrt{\frac{\gamma}{\bar{\gamma}_{FSO}}}\middle|\begin{array}{c}1,\xi^2+1\\ \xi^2,\alpha,\beta,0\end{array}\right)\right)^t \left[1 - \right.$$
$$\left.\left.\frac{\xi^2}{\Gamma(\alpha)\Gamma(\beta)} G_{2,4}^{3,1}\left(\alpha\beta\kappa\sqrt{\frac{\gamma}{\bar{\gamma}_{FSO}}}\middle|\begin{array}{c}1,\xi^2+1\\ \xi^2,\alpha,\beta,0\end{array}\right)\right]\right\}d\gamma.$$

When $t > 1$, because of multiplication of three Meijer-G function and one exponential function, the above integral is unsolvable. In this section exact and asymptotic expression are derived for BER of the proposed system in Gamma-Gamma atmospheric turbulence with the effect of pointing error.

*1) Exact BER*

Substituting the CDF of Gamma-Gamma atmospheric turbulence with the effect of pointing error, from Appendix A into (21), obtains $P_{out}$ of the proposed structure, substituting the result into (24) and using [19], BER of DPSK modulation in Gamma-Gamma atmospheric turbulence with the effect of pointing error equals to:(26)

*2) Asymptotic BER*

Substituting CDF of Gamma-Gamma atmospheric turbulence with the effect of pointing error, from appendix B, into (21), obtains $P_{out}$ of the proposed structure, substituting the result in (24) and using [19], BER of DPSK modulation in

$$P_e \cong$$

$$\begin{cases} \frac{1}{2}\left\{1 + \sum_{k=1}^{N}\sum_{t=0}^{M-1}\sum_{u=0}^{t} \frac{\Omega(\varpi)^t}{\left(1+\frac{k+u}{\bar{\gamma}_{RF}}\right)^{\frac{\beta t}{2}+1}} \left[\Gamma\left(\frac{\beta t}{2}+1\right) - \frac{\xi^2 2^{\alpha+\beta-3}}{\pi\Gamma(\alpha)\Gamma(\beta)} G_{5,8}^{6,3}\left(\frac{(\alpha\beta\kappa)^2}{16\bar{\gamma}_{FSO}\left(1+\frac{k+u}{\bar{\gamma}_{RF}}\right)} \middle| \begin{array}{c} -\frac{\beta t}{2}, 0.5, 1, \frac{\xi^2+1}{2}, \frac{\xi^2+2}{2} \\ \frac{\xi^2}{2}, \frac{\xi^2+1}{2}, \frac{\alpha}{2}, \frac{\alpha+1}{2}, \frac{\beta}{2}, \frac{\beta+1}{2}, 0, \frac{1}{2} \end{array}\right)\right]\right\} & (1) \\[2ex] \frac{1}{2}\left\{1 + \sum_{k=1}^{N}\sum_{t=0}^{M-1}\sum_{u=0}^{t} \frac{\Omega(\rho)^t}{\left(1+\frac{k+u}{\bar{\gamma}_{RF}}\right)^{\frac{\xi^2 t}{2}+1}} \left[\Gamma\left(\frac{\xi^2 t}{2}+1\right) - \frac{\xi^2 2^{\alpha+\beta-3}}{\pi\Gamma(\alpha)\Gamma(\beta)} G_{5,8}^{6,3}\left(\frac{(\alpha\beta\kappa)^2}{16\bar{\gamma}_{FSO}\left(1+\frac{k+u}{\bar{\gamma}_{RF}}\right)} \middle| \begin{array}{c} -\frac{\xi^2 t}{2}, 0.5, 1, \frac{\xi^2+1}{2}, \frac{\xi^2+2}{2} \\ \frac{\xi^2}{2}, \frac{\xi^2+1}{2}, \frac{\alpha}{2}, \frac{\alpha+1}{2}, \frac{\beta}{2}, \frac{\beta+1}{2}, 0, \frac{1}{2} \end{array}\right)\right]\right\} & (2) \\[2ex] \frac{1}{2}\left\{1 + \sum_{k=1}^{N}\sum_{t=0}^{M-1}\sum_{u=0}^{t} \frac{\Omega(\vartheta)^t}{\left(1+\frac{k+u}{\bar{\gamma}_{RF}}\right)^{\frac{\alpha t}{2}+1}} \left[\Gamma\left(\frac{\alpha t}{2}+1\right) - \frac{\xi^2 2^{\alpha+\beta-3}}{\pi\Gamma(\alpha)\Gamma(\beta)} G_{5,8}^{6,3}\left(\frac{(\alpha\beta\kappa)^2}{16\bar{\gamma}_{FSO}\left(1+\frac{k+u}{\bar{\gamma}_{RF}}\right)} \middle| \begin{array}{c} -\frac{\alpha t}{2}, 0.5, 1, \frac{\xi^2+1}{2}, \frac{\xi^2+2}{2} \\ \frac{\xi^2}{2}, \frac{\xi^2+1}{2}, \frac{\alpha}{2}, \frac{\alpha+1}{2}, \frac{\beta}{2}, \frac{\beta+1}{2}, 0, \frac{1}{2} \end{array}\right)\right]\right\} & (3) \end{cases}$$

(27)

Gamma-Gamma atmospheric turbulence with the effect of pointing error becomes equal to:(27)

Substituting (23) into (24) and using [20] BER of DPSK modulation in Negative Exponential atmospheric turbulence becomes equal to:

$$P_e = \frac{1}{2}\left\{1 + \sum_{k=1}^{N}\sum_{t=0}^{M-1}\sum_{u=0}^{t}\sum_{v=0}^{t} \Lambda \frac{1}{1+\frac{k+u}{\bar{\gamma}_{RF}}} \times G_{1,2}^{2,1}\left(\frac{(\lambda(v+1))^2\gamma}{4\bar{\gamma}_{FSO}\left(1+\frac{k+u}{\bar{\gamma}_{RF}}\right)}\middle|\begin{array}{c} - \\ 0, 0.5 \end{array}\right)\right\} \quad (28)$$

### IV. PERFORMANCE EVALUATION OF UNKNOWN CSI AT THE FIRST RELAY SCHEME

According to (6), CDF of $\gamma_{2^{nd}relay}$ is equal to [22]:

$$F_{\gamma_{2^{nd}relay}}(\gamma) = 1 - \int_0^\infty \Pr\left(\gamma_2 \geq \frac{\gamma C}{x}\middle|\gamma_1\right) f_{\gamma_1}(x+\gamma)dx \quad (29)$$

Substituting (8) and (13) into (29) CDF of $\gamma_{2^{nd}relay}$ random variable in Gamma-Gamma atmospheric turbulence with the effect of pointing error becomes equal to:

$$F_{\gamma_{2^{nd}relay}}(\gamma) = 1 - \sum_{k=0}^{N-1}\binom{N-1}{k}(-1)^k \frac{N}{\bar{\gamma}_{RF}} \times \quad (30)$$
$$e^{-\frac{(k+1)\gamma}{\bar{\gamma}_{RF}}} \int_0^\infty e^{-\frac{(k+1)x}{\bar{\gamma}_{RF}}}\left(1 - \frac{\xi^2}{\Gamma(\alpha)\Gamma(\beta)} \times G_{2,4}^{3,1}\left(\alpha\beta\kappa\sqrt{\frac{\gamma C}{x\bar{\gamma}_{FSO}}}\middle|\begin{array}{c} 1, \xi^2+1 \\ \xi^2, \alpha, \beta, 0 \end{array}\right)\right)dx$$

Substituting equivalent Meijer-G form of $G_{2,4}^{3,1}\left(\alpha\beta\kappa\sqrt{\frac{\gamma C}{x\bar{\gamma}_{FSO}}}\middle|\begin{array}{c}1,\xi^2+1\\\xi^2,\alpha,\beta,0\end{array}\right)$ as $G_{4,2}^{1,3}\left(\frac{1}{\alpha\beta\kappa}\sqrt{\frac{x\bar{\gamma}_{FSO}}{\gamma C}}\middle|\begin{array}{c}1-\xi^2,1-\alpha,1-\beta,1\\0,-\xi^2\end{array}\right)$ [20, Eq.07.34.17.0012.01], and using [20, Eq.07.34.21.0088.01] and [20, Eq. 07.34.17.0012.01 ], CDF of $\gamma_{2^{nd}relay}$ random variable in Gamma-Gamma atmospheric turbulence with the effect of pointing error becomes equal to:

$$F_{\gamma_{2^{nd}relay}}(\gamma) = 1 - \sum_{k=0}^{N-1}\binom{N-1}{k}(-1)^k \frac{N}{k+1} \times \quad (31)$$
$$e^{-\frac{(k+1)\gamma}{\bar{\gamma}_{RF}}}\left[1 - \frac{\xi^2 2^{\alpha+\beta-3}}{\pi\Gamma(\alpha)\Gamma(\beta)} G_{4,9}^{7,2}\left(\frac{(\alpha\beta\kappa)^2 C(k+1)\gamma}{16\bar{\gamma}_{FSO}\bar{\gamma}_{RF}}\middle|\begin{array}{c}\psi_1\\\psi_2\end{array}\right)\right]$$

where $\psi_1 = \left\{1, \frac{1}{2}, \frac{\xi^2+2}{2}, \frac{\xi^2+1}{2}\right\}$ and $\psi_2 = \left\{\frac{\xi^2}{2}, \frac{\xi^2+1}{2}, \frac{\alpha}{2}, \frac{\alpha+1}{2}, \frac{\beta}{2}, \frac{\beta+1}{2}, 0, \frac{1}{2}\right\}$.

Substituting (10) and (13) into (29), the CDF of $\gamma_{2^{nd}relay}$ in Negative Exponential atmospheric turbulence becomes equal to:

$$F_{\gamma_{2^{nd}relay}}(\gamma) = 1 - \sum_{k=0}^{N-1}\binom{N-1}{k}(-1)^k \frac{N}{\bar{\gamma}_{RF}} e^{-\frac{(k+1)\gamma}{\bar{\gamma}_{RF}}} \quad (32)$$
$$\times \int_0^\infty e^{-\frac{(k+1)x}{\bar{\gamma}_{RF}}} e^{-\lambda\sqrt{\frac{\gamma C}{x\bar{\gamma}_{FSO}}}} dx.$$

Substituting equivalent Meijer-G form of $e^{-\lambda\sqrt{\frac{\gamma C}{x\bar{\gamma}_{FSO}}}}$ as $\frac{1}{\sqrt{\pi}} G_{2,0}^{0,2}\left(\frac{4x\bar{\gamma}_{FSO}}{\lambda^2\gamma C}\middle|\begin{array}{c}1,0.5\\-\end{array}\right)$ [19], and using [19] and [19], CDF of $\gamma_{2^{nd}relay}$ in Negative Exponential atmospheric turbulence becomes equal to:

$$F_{\gamma_{2^{nd}relay}}(\gamma) = 1 - \sum_{k=0}^{N-1}\binom{N-1}{k}(-1)^k \times \quad (33)$$
$$\frac{N}{(k+1)} e^{-\frac{(k+1)\gamma}{\bar{\gamma}_{RF}}} \frac{1}{\sqrt{\pi}} G_{0,3}^{3,0}\left(\frac{\lambda^2\gamma C(k+1)}{4\bar{\gamma}_{FSO}\bar{\gamma}_{RF}}\middle|\begin{array}{c}-\\1,0,0.5\end{array}\right).$$

#### A. Outage Probability

According to (14), after substituting (31), (8) and (9) into (19) and substituting binomial expansion of $\left[1 - \frac{\xi^2}{\Gamma(\alpha)\Gamma(\beta)}\left(1 - e^{-\frac{\gamma_{th}}{\bar{\gamma}_{RF}}}\right) G_{2,4}^{3,1}\left(\alpha\beta\kappa\sqrt{\frac{\gamma_{th}}{\bar{\gamma}_{FSO}}}\middle|\begin{array}{c}1,\xi^2+1\\\xi^2,\alpha,\beta,0\end{array}\right)\right]^{M-1}$, $P_{out}$ of the proposed structure in Gamma-Gamma atmospheric turbulence with the effect of pointing error becomes equal to:

$$P_{out}(\gamma_{th}) = 1 - \sum_{k=0}^{N-1}\sum_{t=0}^{M-1}\sum_{u=0}^{t} \varsigma e^{-\frac{(k+u+1)\gamma_{th}}{\bar{\gamma}_{RF}}} \times \quad (34)$$
$$\left(\frac{\xi^2}{\Gamma(\alpha)\Gamma(\beta)} G_{2,4}^{3,1}\left(\alpha\beta\kappa\sqrt{\frac{\gamma_{th}}{\bar{\gamma}_{FSO}}}\middle|\begin{array}{c}1,\xi^2+1\\\xi^2,\alpha,\beta,0\end{array}\right)\right)^t \left[1 - \frac{\xi^2 2^{\alpha+\beta-3}}{\pi\Gamma(\alpha)\Gamma(\beta)} G_{4,9}^{7,2}\left(\frac{(\alpha\beta\kappa)^2 C(k+1)\gamma_{th}}{16\bar{\gamma}_{FSO}\bar{\gamma}_{RF}}\middle|\begin{array}{c}\psi_1\\\psi_2\end{array}\right)\right],$$

$$P_e = \frac{1}{2}\left\{1 + \sum_{k=0}^{N-1}\sum_{t=0}^{M-1}\sum_{u=0}^{t}\sum_{k_1=0}^{t}\sum_{k_2=0}^{k_1}\sum_{n=0}^{\infty} \varsigma \binom{t}{k_1}\binom{k_1}{k_2} X_0^{t-k_1}\left(Y_n^{(k_1-k_2)} * Z_n^{(k_2)}\right) \frac{\left(\frac{1}{\bar{\gamma}_{FSO}}\right)^{\frac{n+\xi^2(t-k_1)+\alpha(k_1-k_2)+\beta k_2}{2}}}{\left(1+\frac{k+u+1}{\bar{\gamma}_{RF}}\right)^{1+\frac{n+\xi^2(t-k_1)+\alpha(k_1-k_2)+\beta k_2}{2}}} \times \right.$$

$$\left.\left[\Gamma\left(1+\frac{n+\xi^2(t-k_1)+\alpha(k_1-k_2)+\beta k_2}{2}\right) - \frac{\xi^2 2^{\alpha+\beta-3}}{\pi\Gamma(\alpha)\Gamma(\beta)} G_{5,9}^{7,3}\left(\frac{(\alpha\beta\kappa)^2 C(k+1)}{16\bar{\gamma}_{FSO}(\bar{\gamma}_{RF}+k+u+1)}\Bigg|\begin{matrix}-\frac{n+\xi^2(t-k_1)+\alpha(k_1-k_2)+\beta k_2}{2},\psi_1\\ \psi_2\end{matrix}\right)\right]\right\}$$
(38)

where $\varsigma = \binom{N-1}{k}\binom{M-1}{t}\binom{t}{u}(-1)^{k+t+u}\frac{N}{k+1}$.

Substituting (33), (9) and (10) into (19), $P_{out}$ of the proposed structure in Negative Exponential atmospheric turbulence becomes equal to:

$$P_{out}(\gamma_{th}) = 1 - \sum_{k=0}^{N-1}\binom{N-1}{k}(-1)^k \frac{N}{(k+1)} \times \tag{35}$$
$$e^{-\frac{(k+1)\gamma_{th}}{\bar{\gamma}_{RF}}} \frac{1}{\sqrt{\pi}} G_{0,3}^{3,0}\left(\frac{\lambda^2 \gamma_{th} C(k+1)}{4\bar{\gamma}_{FSO}\bar{\gamma}_{RF}}\Bigg|\begin{matrix}-\\1,0,0.5\end{matrix}\right)\left[1-\left(1-e^{-\frac{\gamma_{th}}{\bar{\gamma}_{RF}}}\right)\left(1-e^{-\lambda\sqrt{\frac{\gamma_{th}}{\bar{\gamma}_{FSO}}}}\right)\right]^{M-1}.$$

Substituting binomial expansion of $\left[1-\left(1-e^{-\frac{\gamma_{th}}{\bar{\gamma}_{RF}}}\right)\left(1-e^{-\lambda\sqrt{\frac{\gamma_{th}}{\bar{\gamma}_{FSO}}}}\right)\right]^{M-1}$ as $\sum_{t=0}^{M-1}\sum_{u=0}^{t}\sum_{v=0}^{t}\binom{M-1}{t}\binom{t}{u}\binom{t}{v} \times (-1)^{t+u+v} e^{-\frac{u\gamma_{th}}{\bar{\gamma}_{RF}}} e^{-\lambda v\sqrt{\frac{\gamma_{th}}{\bar{\gamma}_{FSO}}}}$, and substituting equivalent Meijer-G form of $e^{-\lambda v\sqrt{\frac{\gamma_{th}}{\bar{\gamma}_{FSO}}}}$ as $\frac{1}{\sqrt{\pi}} G_{0,2}^{2,0}\left(\frac{(\lambda v)^2 \gamma_{th}}{4\bar{\gamma}_{FSO}}\Bigg|\begin{matrix}-\\0,0.5\end{matrix}\right)$, $P_{out}$ of the proposed structure in Negative Exponential atmospheric turbulence becomes equal to:

$$P_{out}(\gamma_{th}) = 1 - \sum_{k=0}^{N-1}\sum_{t=0}^{M-1}\sum_{u=0}^{t}\sum_{v=0}^{t} \varrho\, e^{-\frac{(k+u+1)\gamma_{th}}{\bar{\gamma}_{RF}}} \tag{36}$$
$$\times G_{0,2}^{2,0}\left(\frac{(\lambda v)^2 \gamma_{th}}{4\bar{\gamma}_{FSO}}\Bigg|\begin{matrix}-\\0,0.5\end{matrix}\right)$$
$$\times G_{0,3}^{3,0}\left(\frac{\lambda^2 C(k+1)\gamma_{th}}{4\bar{\gamma}_{FSO}\bar{\gamma}_{RF}}\Bigg|\begin{matrix}-\\1,0,0.5\end{matrix}\right)$$

where, $\varrho = \binom{N-1}{k}\binom{M-1}{t}\binom{t}{u}\binom{t}{v}(-1)^{k+t+u+v}\frac{N}{\pi(k+1)}$.

### B. Bit Error Rate

Substituting (34) into (24), BER of DPSK modulation in Gamma-Gamma atmospheric turbulence with the effect of pointing error is equal to:

$$P_e = \frac{1}{2}\int_0^\infty e^{-\gamma}\left\{1 - \sum_{k=0}^{N-1}\sum_{t=0}^{M-1}\sum_{u=0}^{t} \varsigma e^{-\frac{(k+u+1)\gamma}{\bar{\gamma}_{RF}}} \times \right.\tag{37}$$
$$\left.\left(\frac{\xi^2}{\Gamma(\alpha)\Gamma(\beta)} G_{2,4}^{3,1}\left(\alpha\beta\kappa\sqrt{\frac{\gamma}{\bar{\gamma}_{FSO}}}\Bigg|\begin{matrix}1,\xi^2+1\\ \xi^2,\alpha,\beta,0\end{matrix}\right)\right)^t \times \left[1 - \frac{\xi^2 2^{\alpha+\beta-3}}{\pi\Gamma(\alpha)\Gamma(\beta)} G_{4,9}^{7,2}\left(\frac{(\alpha\beta\kappa)^2 C(k+1)\gamma}{16\bar{\gamma}_{FSO}\bar{\gamma}_{RF}}\Bigg|\begin{matrix}\psi_1\\ \psi_2\end{matrix}\right)\right]\right\} d\gamma.$$

When $t > 1$, because of multiplication of three Meijer-G function and one exponential function, the above integral is unsolvable. In this section exact and asymptotic expressions are derived for BER of the proposed system in Gamma-Gamma atmospheric turbulence with the effect of pointing error.

*1) Exact BER*

Substituting CDF of Gamma-Gamma atmospheric turbulence with the effect of pointing error from appendix A, into (34), obtains new form of $P_{out}$ of the proposed structure; substituting the result into (24) and using [19], BER of DPSK modulation in Gamma-Gamma atmospheric turbulence with the effect of pointing error is equal to:(38)

*2) Asymptotic BER*

Substituting CDF of Gamma-Gamma atmospheric turbulence with the effect of pointing error, from Appendix B, into (34), obtains $P_{out}$ of the proposed structure, substituting

$$P_e \cong \begin{cases} \frac{1}{2}\left\{1 - \sum_{k=0}^{N-1}\sum_{t=0}^{M-1}\sum_{u=0}^{t} \frac{\varsigma(\varpi)^t}{\left(1+\frac{k+u+1}{\bar{\gamma}_{RF}}\right)^{\frac{\beta t}{2}+1}}\left[\Gamma\left(\frac{\beta t}{2}+1\right) - \frac{\xi^2 2^{\alpha+\beta-3}}{\pi\Gamma(\alpha)\Gamma(\beta)} G_{5,9}^{7,3}\left(\frac{(\alpha\beta\kappa)^2 C(k+1)}{16\bar{\gamma}_{FSO}(\bar{\gamma}_{RF}+k+u+1)}\Bigg|\begin{matrix}-\frac{\beta t}{2},\psi_1\\ \psi_2\end{matrix}\right)\right]\right\} & (1)\\[2mm]
\frac{1}{2}\left\{1 - \sum_{k=0}^{N-1}\sum_{t=0}^{M-1}\sum_{u=0}^{t} \frac{\varsigma(\rho)^t}{\left(1+\frac{k+u+1}{\bar{\gamma}_{RF}}\right)^{\frac{\xi^2 t}{2}+1}}\left[\Gamma\left(\frac{\xi^2 t}{2}+1\right) - \frac{\xi^2 2^{\alpha+\beta-3}}{\pi\Gamma(\alpha)\Gamma(\beta)} G_{5,9}^{7,3}\left(\frac{(\alpha\beta\kappa)^2 C(k+1)}{16\bar{\gamma}_{FSO}(\bar{\gamma}_{RF}+k+u+1)}\Bigg|\begin{matrix}-\frac{\xi^2 t}{2},\psi_1\\ \psi_2\end{matrix}\right)\right]\right\} & (2)\\[2mm]
\frac{1}{2}\left\{1 - \sum_{k=0}^{N-1}\sum_{t=0}^{M-1}\sum_{u=0}^{t} \frac{\varsigma(\vartheta)^t}{\left(1+\frac{k+u+1}{\bar{\gamma}_{RF}}\right)^{\frac{\alpha t}{2}+1}}\left[\Gamma\left(\frac{\alpha t}{2}+1\right) - \frac{\xi^2 2^{\alpha+\beta-3}}{\pi\Gamma(\alpha)\Gamma(\beta)} G_{5,9}^{7,3}\left(\frac{(\alpha\beta\kappa)^2 C(k+1)}{16\bar{\gamma}_{FSO}(\bar{\gamma}_{RF}+k+u+1)}\Bigg|\begin{matrix}-\frac{\alpha t}{2},\psi_1\\ \psi_2\end{matrix}\right)\right]\right\} & (3)
\end{cases} \quad (39)$$

the result into (24) and using [19], BER of DPSK modulation in Gamma-Gamma atmospheric turbulence with the effect of pointing error becomes equal to:(39)

Substituting (36) into (24), and using [20] BER of DPSK modulation in Negative exponential atmospheric turbulence is equal to:

$$P_e = \frac{1}{2}\left\{1 - \sum_{k=0}^{N-1}\sum_{t=0}^{M-1}\sum_{u=0}^{t}\sum_{v=0}^{t} Q \frac{1}{1+\frac{k+u+1}{\bar{\gamma}_{RF}}} \times G_{1,0:0,2:0,3}^{1,0:2,0:3,0}\left(1 \left| \begin{matrix} - \\ - \end{matrix} \right| \begin{matrix} - \\ 0,0.5 \end{matrix} \left| \begin{matrix} - \\ 1,0,0.5 \end{matrix} \right| \frac{(\lambda v)^2}{4\bar{\gamma}_{FSO}\left(1+\frac{k+u+1}{\bar{\gamma}_{RF}}\right)}, \frac{\lambda^2 C(k+1)}{4\bar{\gamma}_{FSO}(\bar{\gamma}_{RF}+k+u+1)}\right)\right\},$$
(40)

where

$G_{p_1,q_1:p_2,q_2:p_3,q_3}^{n_1,m_1:n_2,m_2:n_3,m_3}\left(\begin{matrix}a_1,a_2,\ldots,a_{p_1}\\b_1,b_2,\ldots,b_{q_1}\end{matrix}\left|\begin{matrix}c_1,c_2,\ldots,c_{p_2}\\d_1,d_2,\ldots,d_{q_2}\end{matrix}\right|\begin{matrix}e_1,e_2,\ldots,e_{p_3}\\f_1,f_2,\ldots,f_{q_3}\end{matrix}\right|x,y\right)$ is the Extended Bivariate Meijer-G function [24].

## V. SIMULATION RESULTS

In this section, exact and asymptotic analytical results are compared with MATLAB simulations. Impact of number of users and relays on the performance of the proposed structure is investigated. Also moderate ($\alpha = 4, \beta = 1.9, \xi = 10.45$) and strong ($\alpha = 4.2, \beta = 1.4, \xi = 2.45$) regimes are considered for of Gamma-Gamma atmospheric turbulence with the effect of pointing error. $M$ and $N$ are number of relays and users, respectively. $\gamma_{th}$ is outage threshold SNR of the proposed system. It is assumed that FSO and RF links, have equal average SNR ($\bar{\gamma}_{FSO} = \bar{\gamma}_{RF} = \gamma_{avg}$), $\eta = 1$ and $C = 1$.

In Fig. 2, Outage Probability of the proposed structure is plotted as a function of average SNR for various number of relays, for both cases of known CSI and unknown CSI at the first relay, for moderate regime of Gamma-Gamma atmospheric turbulence with the effect of pointing error, when number of users is $N = 2$ and $\gamma_{th} = 10dB$. As can be seen, in the case of known CSI, at low $\gamma_{avg}$, there is little performance difference between various numbers of relays, but at high $\gamma_{avg}$ this difference evaporates. Since relay addition increases system capacity. The proposed structure, increases number of users that can be served, whereas maintaining performance of the system. This feature, is especially useful for modern communication systems, which should service huge number of users with favorable performance. In the case of unknown CSI, the proposed structure is dependent on the number of relays. But in the second situation, amplification gain is fixed at all circumstances, thereby system is sensitive to changes. At low target $P_{out}$, $\gamma_{avg}$ difference between various numbers of relay is more. For example, at $P_{out} = 10^{-2}$ the difference between modes of $M = 1$ and $M = 2$ or between $M = 1$ and $M = 3$ is $2dB$ and $4dB$, respectively. At $P_{out} = 10^{-4}$, these differences become $2.5dB$ and $5dB$, respectively.

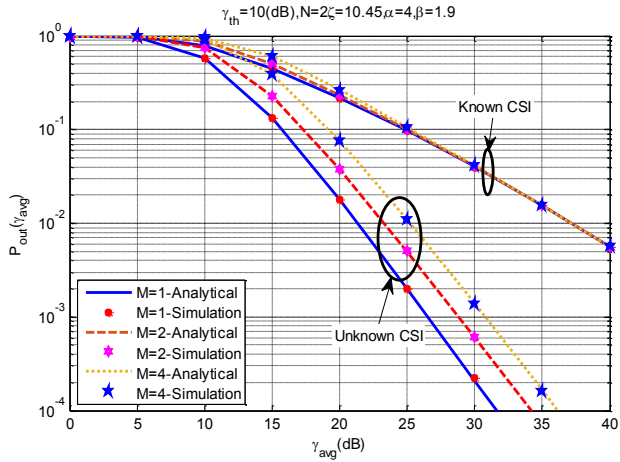

Fig. 2. Outage Probability of the proposed structure as a function of average SNR for various number of relays, for both cases of known CSI and unknown CSI at the first relay, for moderate regime of Gamma-Gamma atmospheric turbulence with the effect of pointing error, when number of

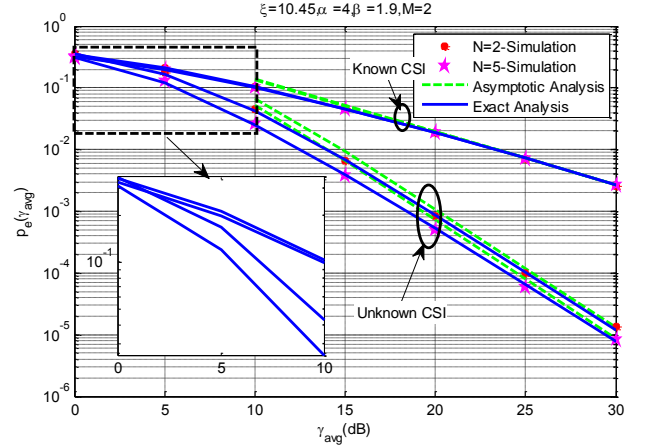

Fig. 3. Bit Error Rate of the proposed structure as a function of average SNR for various number of users, for moderate regime of Gamma-Gamma atmospheric turbulence with the effect of pointing error for both cases of known CSI and unknown CSI, when number of relays is $M = 2$.

In Fig. 3, Bit Error Rate of the proposed structure is plotted as a function of average SNR for various number of users, for moderate regime of Gamma-Gamma atmospheric turbulence with the effect of pointing error, for both cases of known CSI and unknown CSI, when number of relays is $M = 2$. As can be seen, in both cases of known CSI and unknown CSI, system performance has very low dependence on the number of users within the cell. This difference decreases by increase in $\gamma_{avg}$. Of course in the case of unknown CSI, system is more dependent on the number of users within the cell. Because when CSI is available, relay adjusts the amplification gain based on conditions, but in the other case, the signal is amplified with fixed gain. It is an advantage of the proposed structure that does not need to adapt its parameters according to channel conditions. Although the case of known CSI has more complexity, but it offers a reliable data transmission. In the case of unknown CSI, system performance compared with the case of known CSI, is more dependent on the number of users within the cell. At high $\gamma_{avg}$, system sensitivity to the number of users reduces, in the sense that at high $\gamma_{avg}$ system performs almost independent from number of users. Therefore, the proposed system is suitable for networks with frequent changes in population. Generally, increasing number of users improves the performance of the proposed system for both cases of known CSI and unknown CSI, because the user with the highest $\gamma_{avg}$

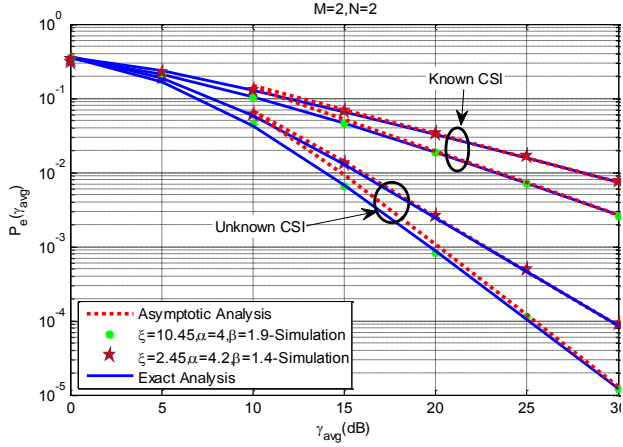

Fig. 4. Bit Error Rate of the proposed structure as a function of average SNR, for moderate and strong regimes of Gamma-Gamma atmospheric turbulence with the effect of pointing error for both cases of known CSI and unknown CSI, when number of relays is $M = 2$ and number of users is $N = 2$.

is selected for communication.

In Fig. 4, Bit Error Rate of the proposed structure is plotted as a function of average SNR, for moderate and strong regimes of Gamma-Gamma atmospheric turbulence with the effect of pointing error for both cases of known CSI and unknown CSI, when number of relays is $M = 2$ and number of users is $N = 2$. The results indicate that in the case of known CSI and unknown CSI, at different target $P_e$, $\gamma_{avg}$ difference between system performance at moderate and strong atmospheric turbulence regimes is variable. In the sense that at these atmospheric turbulence regimes, the system has different gains and it depends on target $P_e$, the differences increases when the effect of atmospheric turbulence increases and the effect of noise decreases.

In Fig. 5, Outage Probability of the proposed structure is plotted as a function of average SNR for various number of users, for both cases of known CSI and unknown CSI, for Negative Exponential atmospheric turbulence with unit variance, when number of relay is $M = 2$ and $\gamma_{th} = 10dB$. As can be seen, in the case of known CSI, performance of the proposed structure is independent of the cell population and in the case of unknown CSI, the proposed structure has slight dependence on the number of users within the cell. The more number of users, the more likely can find a signal with desirable $\gamma_{avg}$ and therefore the system performs better. The results are the same as the case of Gamma-Gamma atmospheric turbulence except that $\gamma_{avg}$ difference between different modes is reduced.

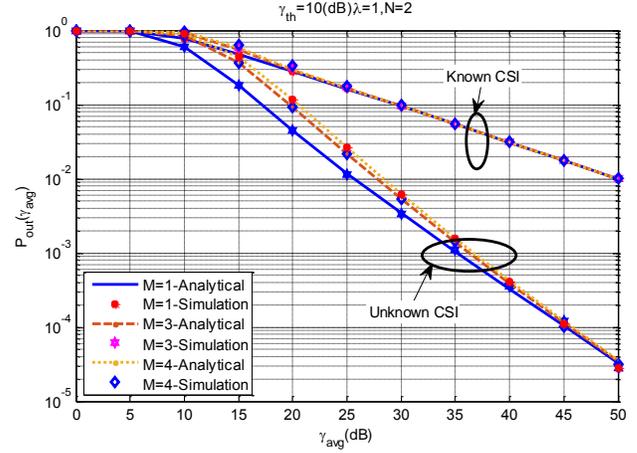

Fig. 6. Outage Probability of the proposed structure as a function of average SNR for various number of relays for both cases of known CSI and unknown CSI, for Negative Exponential atmospheric turbulence with unit variance, when number of users is $N = 2$.

In Fig. 6, Outage Probability of the proposed structure is plotted as a function of average SNR for various number of relays for both cases of known CSI and unknown CSI, for Negative Exponential atmospheric turbulence with unit variance, when number of users is $N = 2$. As can be seen, in both cases of known CSI and unknown CSI, at low $\gamma_{avg}$, there is slight performance difference between systems with different number of relays, and this difference is a bit more in the case of unknown CSI. In both cases of known CSI and unknown CSI, at high $\gamma_{avg}$, system performance becomes independent of the number of relays. Generally, series relay structure, degrades system performance, because of frequent

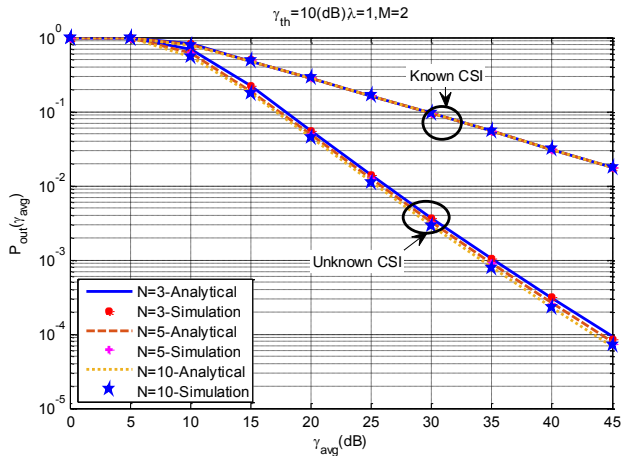

Fig. 5. Outage Probability of the proposed structure as a function of average SNR for various number of users, for both cases of known CSI and unknown CSI, for Negative Exponential atmospheric turbulence with unit variance, when number of relay is $M = 2$ and $\gamma_{th} = 10dB$.

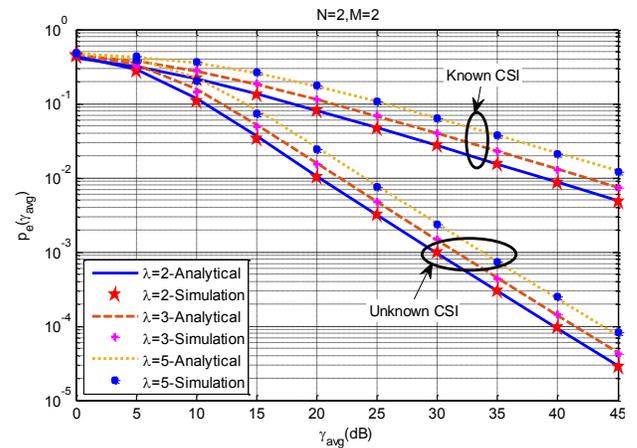

Fig. 7. Bit Error Rate of the proposed structure is plotted as a function of average SNR for various variances of Negative Exponential atmospheric turbulence for both cases of known CSI and unknown CSI, when number of relays is $M = 2$ and number of users is $N = 2$.

$$f_{\gamma_{FSO}}(\gamma) = \frac{\xi^2 \Gamma(\alpha-\xi^2)\Gamma(\beta-\xi^2)}{2\Gamma(\alpha)\Gamma(\beta)\gamma}\left(\alpha\beta\kappa\sqrt{\frac{\gamma}{\bar{\gamma}_{FSO}}}\right)^{\xi^2} {}_1F_2\left(0;1-\alpha+\xi^2,1-\beta+\xi^2;\alpha\beta\kappa\sqrt{\frac{\gamma}{\bar{\gamma}_{FSO}}}\right) + \frac{\xi^2\Gamma(\xi^2-\alpha)\Gamma(\beta-\alpha)}{2\Gamma(\alpha)\Gamma(\beta)\Gamma(\xi^2+1-\alpha)\gamma} \times$$
$$\left(\alpha\beta\kappa\sqrt{\frac{\gamma}{\bar{\gamma}_{FSO}}}\right)^{\alpha} \times {}_1F_2\left(\alpha-\xi^2;1-\xi^2+\alpha,1-\beta+\alpha;\alpha\beta\kappa\sqrt{\frac{\gamma}{\bar{\gamma}_{FSO}}}\right) + \frac{\xi^2\Gamma(\alpha-\beta)\Gamma(\xi^2-\beta)}{2\Gamma(\alpha)\Gamma(\beta)\Gamma(\xi^2+1-\beta)\gamma}\left(\alpha\beta\kappa\sqrt{\frac{\gamma}{\bar{\gamma}_{FSO}}}\right)^{\beta} \times$$
$${}_1F_2\left(\beta-\xi^2;1-\xi^2+\beta,1-\alpha+\beta;\alpha\beta\kappa\sqrt{\frac{\gamma}{\bar{\gamma}_{FSO}}}\right) \quad (41)$$

$$f_{\gamma_{FSO}}(\gamma) = \frac{\xi^2\Gamma(\alpha-\xi^2)\Gamma(\beta-\xi^2)}{2\Gamma(\alpha)\Gamma(\beta)}(\alpha\beta\kappa)^{\xi^2}\left(\frac{\gamma}{\bar{\gamma}_{FSO}}\right)^{\frac{\xi^2}{2}-1} + \sum_{n=0}^{\infty}\frac{\xi^2\Gamma(\xi^2-\alpha)\Gamma(\beta-\alpha)(\alpha-\xi^2)_n}{2\Gamma(\alpha)\Gamma(\beta)\Gamma(\xi^2+1-\alpha)(1-\xi^2+\alpha)_n(1-\beta+\alpha)_n n!} \times$$
$$(\alpha\beta\kappa)^{n+\alpha}\left(\frac{\gamma}{\bar{\gamma}_{FSO}}\right)^{\frac{n+\alpha}{2}-1} + \sum_{n=0}^{\infty}\frac{\xi^2\Gamma(\alpha-\beta)\Gamma(\xi^2-\beta)(\beta-\xi^2)_n}{2\Gamma(\alpha)\Gamma(\beta)\Gamma(\xi^2+1-\beta)(1-\xi^2+\beta)_n(1-\alpha+\beta)_n}(\alpha\beta\kappa)^{n+\beta}\left(\frac{\gamma}{\bar{\gamma}_{FSO}}\right)^{\frac{n+\beta}{2}-1} \quad (42)$$

$$F_{\gamma_{FSO}}(\gamma) = \frac{\Gamma(\alpha-\xi^2)\Gamma(\beta-\xi^2)}{\Gamma(\alpha)\Gamma(\beta)}(\alpha\beta\kappa)^{\xi^2}\left(\frac{\gamma}{\bar{\gamma}_{FSO}}\right)^{\frac{\xi^2}{2}} + \sum_{n=0}^{\infty}\frac{\xi^2\Gamma(\xi^2-\alpha)\Gamma(\beta-\alpha)(\alpha-\xi^2)_n}{(n+\alpha)\Gamma(\alpha)\Gamma(\beta)\Gamma(\xi^2+1-\alpha)(1-\xi^2+\alpha)_n(1-\beta+\alpha)_n n!}(\alpha\beta\kappa)^{n+\alpha}\left(\frac{\gamma}{\bar{\gamma}_{FSO}}\right)^{\frac{n+\alpha}{2}} +$$
$$\sum_{n=0}^{\infty}\frac{\xi^2\Gamma(\alpha-\beta)\Gamma(\xi^2-\beta)(\beta-\xi^2)_n}{(n+\beta)\Gamma(\alpha)\Gamma(\beta)\Gamma(\xi^2+1-\beta)(1-\xi^2+\beta)_n(1-\alpha+\beta)_n}(\alpha\beta\kappa)^{n+\beta}\left(\frac{\gamma}{\bar{\gamma}_{FSO}}\right)^{\frac{n+\beta}{2}} = X_0\left(\frac{\gamma}{\bar{\gamma}_{FSO}}\right)^{\frac{\xi^2}{2}} + \sum_{n=0}^{\infty}Y_n\left(\frac{\gamma}{\bar{\gamma}_{FSO}}\right)^{\frac{n+\alpha}{2}} + \sum_{n=0}^{\infty}Z_n\left(\frac{\gamma}{\bar{\gamma}_{FSO}}\right)^{\frac{n+\beta}{2}}$$
$$(43)$$

decisions made by relays. In fact, low dependence on the number of relays, increases capacity whereas maintaining system performance, the same as in the case of Gamma-Gamma atmospheric turbulence.

In Fig. 7, Bit Error Rate of the proposed structure is plotted as a function of average SNR for various variances of Negative Exponential atmospheric turbulence for both cases of known CSI and unknown CSI, when number of relays is $M = 2$ and number of users is $N = 2$. As can be seen, in case of known CSI, $\gamma_{avg}$ difference between the two atmospheric turbulence with variances of $\lambda = 1, \lambda = 2$ is about 5dB and between $\lambda = 2, \lambda = 5$ is about 7dB, and in the case of unknown CSI, these differences are respectively about 1.5dB and 2dB. This difference is because when CSI is unknown, the amplification gain is chosen fixed and manually, usually operators define this gain according to the worst case scenario and that is exactly why this scheme performs better than the known CSI scheme.

## APPENDIX A

Using [20,Eq.07.34.26.0004.01], the pdf of Gamma-Gamma atmospheric turbulence with the effect of pointing error becomes equal to (41), where ${}_pF_q(a_1,\dots,a_p;b_1,\dots,b_q;z)$ is Hyper-geometric function [20.Eq, 07.23.02.0001.01]. Using [20,Eq. 07.23.02.0001.01], the above expression becomes equal to (42), where $(.)_n$ is the pochhammer symbol [22]. Integrating (42), CDF of Gamma-Gamma atmospheric turbulence with the effect of pointing error becomes equal to (43). Substituting trinomial expansion of $\left(X_0\left(\frac{\gamma_{th}}{\bar{\gamma}_{FSO}}\right)^{\frac{\xi^2}{2}} + \sum_{n=0}^{\infty}Y_n\left(\frac{\gamma_{th}}{\bar{\gamma}_{FSO}}\right)^{\frac{n+\alpha}{2}} + \sum_{n=0}^{\infty}Z_n\left(\frac{\gamma_{th}}{\bar{\gamma}_{FSO}}\right)^{\frac{n+\beta}{2}}\right)^t$ as $\sum_{k_1=0}^{t}\sum_{k_2=0}^{k_1}\times \binom{t}{k_1}\binom{k_1}{k_2}\left(X_0\left(\frac{\gamma_{th}}{\bar{\gamma}_{FSO}}\right)^{\frac{\xi^2}{2}}\right)^{t-k_1}\left(\sum_{n=0}^{\infty}Y_n\left(\frac{\gamma_{th}}{\bar{\gamma}_{FSO}}\right)^{\frac{n+\alpha}{2}}\right)^{k_1-k_2} \times \left(\sum_{n=0}^{\infty}Z_n\left(\frac{\gamma_{th}}{\bar{\gamma}_{FSO}}\right)^{\frac{n+\beta}{2}}\right)^{k_2}$ and after some mathematical simplification, the CDF of Gamma-Gamma atmospheric turbulence with the effect of pointing error becomes as follows:

$$F_{\gamma_{FSO}}(\gamma) = \sum_{k_1=0}^{t}\sum_{k_2=0}^{k_1}\sum_{n=0}^{\infty}\binom{t}{k_1}\binom{k_1}{k_2}X_0^{t-k_1} \times \left(Y_n^{(k_1-k_2)} * Z_n^{(k_2)}\right)\left(\frac{\gamma_{th}}{\bar{\gamma}_{FSO}}\right)^{\frac{n+\xi^2(t-k_1)+\alpha(k_1-k_2)+\beta k_2}{2}} \quad (44)$$

Where $*$ denotes the convolution and the subscript $h_n^{(k)}$ means that $h_n$ is convolved k-1 times with itself.

## APPENDIX B

Using approximation of [20,Eq. 07.34.06.0006.01], the pdf of Gamma-Gamma atmospheric turbulence with the effect of pointing error becomes equal to:

$$f_{\gamma_{FSO}}(\gamma) = \qquad (45)$$
$$\begin{cases} \frac{\xi^2\Gamma(\alpha-\beta)\Gamma(\xi^2-\beta)}{2\Gamma(\alpha)\Gamma(\beta)\Gamma(\xi^2+1-\beta)\gamma}\left(\alpha\beta\kappa\sqrt{\frac{\gamma}{\bar{\gamma}_{FSO}}}\right)^{\beta} & \xi^2 > \beta, \alpha > \beta \\ \frac{\xi^2\Gamma(\beta-\xi^2)\Gamma(\alpha-\xi^2)}{2\Gamma(\alpha)\Gamma(\beta)\gamma}\left(\alpha\beta\kappa\sqrt{\frac{\gamma}{\bar{\gamma}_{FSO}}}\right)^{\xi^2} & \alpha > \xi^2, \beta > \xi^2 \\ \frac{\xi^2\Gamma(\xi^2-\alpha)\Gamma(\beta-\alpha)}{2\Gamma(\alpha)\Gamma(\beta)\Gamma(\xi^2+1-\alpha)\gamma}\left(\alpha\beta\kappa\sqrt{\frac{\gamma}{\bar{\gamma}_{FSO}}}\right)^{\alpha} & \beta > \alpha, \xi^2 > \alpha \end{cases}$$

Integrating the above equation, the CDF of Gamma-Gamma atmospheric turbulence with the effect of pointing error becomes as follows:

$$F_{\gamma_{FSO}}(\gamma) = \qquad (46)$$
$$\begin{cases} \frac{\xi^2\Gamma(\alpha-\beta)}{\Gamma(\alpha)\Gamma(\beta+1)\Gamma(\xi^2-\beta)}\left(\alpha\beta\kappa\sqrt{\frac{1}{\bar{\gamma}_{FSO}}}\right)^{\beta}\gamma^{\beta/2} & \xi^2 > \beta, \alpha > \beta \\ \frac{\Gamma(\beta-\xi^2)\Gamma(\alpha-\xi^2)}{\Gamma(\alpha)\Gamma(\beta)}\left(\alpha\beta\kappa\sqrt{\frac{1}{\bar{\gamma}_{FSO}}}\right)^{\xi^2}\gamma^{\xi^2/2} & \alpha > \xi^2, \beta > \xi^2 \\ \frac{\xi^2\Gamma(\beta-\alpha)}{\Gamma(\alpha+1)\Gamma(\beta)\Gamma(\xi^2-\alpha)}\left(\alpha\beta\kappa\sqrt{\frac{1}{\bar{\gamma}_{FSO}}}\right)^{\alpha}\gamma^{\alpha/2} & \beta > \alpha, \xi^2 > \alpha \end{cases}$$

$$= \begin{cases} \varpi \gamma^{\beta/2} & (1) \\ \rho \gamma^{\xi^2/2} & (2) \\ \vartheta \gamma^{\alpha/2} & (3) \end{cases}$$

## VI. Conclusion

In this paper, a novel multi-hop relay-assisted hybrid FSO / RF communication system is presented. In this system a relay connects RF users to the source Base Station and a multi-hop hybrid FSO / RF link connects source and destination Base Stations. This structure is suitable where direct RF connection between mobile user and the source Base Station is not possible. FSO link in moderate to strong atmospheric turbulences has Gamma-Gamma distribution with the effect of pointing error and in saturate atmospheric turbulence has Negative Exponential atmospheric turbulence. Also RF link has Rayleigh fading. Finally, the proposed structure is investigated at various number of users and relays. The proposed structure is investigated for both cases of known CSI and unknown CSI at the first relay. Results indicate that the proposed system has low dependence on the number of users, therefore the proposed structure is suitable for areas in which population density changes frequently. Also the proposed structure, at Negative exponential atmospheric turbulence has small dependence on the number of relays, but this dependence is a bit more for Gamma-Gamma atmospheric turbulence. Therefore, the proposed structure increases capacity whereas maintaining performance of the system.

**M. A. Amirabadi** was born in Zahedan, Iran, in 1993. He received the B.Sc. degree in Optics & Laser Engineering from Malek-e-Ashtar University of Technology, Isfahan, Iran, in 2015. Now he is studying M.Sc. in Communication Engineering in Iran University of Science and Technology, Tehran, Iran. He worked with Shaheed Chamran Research Center, Isfahan, as a Research Engineer, from 2011 to 2015. His research interests include Free Space Optical Communications, Relay-Assisted Networks, and Hybrid FSO/RF systems.